# The composite theory as the explanation of Haldane's rule should be abandoned

Ren-Xue Wang [*]

BC Cancer Research Centre, BC Cancer Agency, Vancouver, British Columbia, Canada

BC Cancer Research Centre, BC Cancer Agency, 675 West 10th Avenue, Vancouver, British Columbia, Canada

[*] Tel.: 604 675 8000 Ext. 7709; Fax: 604 675 8185; E-mail address: rwang@bccrc.ca

16  **Abstract**

17      In 1922, JBS Haldane discovered an intriguing bias of postzygotic isolation

18  during early speciation: the heterogametic sex of $F_1$ hybrids between closely related

19  species or subspecies is more susceptible to sterility or inviability than the homogametic

20  sex. This phenomenon, now known as Haldane's rule, has been repeatedly confirmed

21  across broad taxa in diecious animals and plants. Currently, the dominant view in the

22  field of speciation genetics believes that Haldane's rule for sterility, inviability, male

23  heterogamety and female heterogametic belongs to different entities; and Haldane's rule

24  in these subdivisions has different causes, which operate coincidentally and/or

25  collectively resulting in this striking bias against the heterogametic sex in hybridization.

26  This view, known as the composite theory, was developed after many unsuccessful quests

27  in searching for a unitary genetic mechanism. The composite theory has multiple sub-

28  theories. The dominance theory and the faster male theory are the major ones. In this

29  note, I challenge the composite theory and its scientific validity. By declaring Haldane's

30  rule as a composite phenomenon caused by multiple mechanisms

31  coincidentally/collectively, the composite theory becomes a self-fulfilling prophecy and

32  untestable. I believe that the composite theory is an *ad hoc* hypothesis that lacks

33  falsifiability, refutability and testability that a scientific theory requires. It is my belief

34  that the composite theory does not provide meaningful insights for the study of speciation

35  and should be abandoned.

36

37        How many times in the history of science has a seemingly correct theory been

38    falsified based on evidence?

39        How many times then, has such a theory later been revived and announced correct

40    again, while the falsifying evidence still stands?

41        How many times has one single natural phenomenon been explained by many

42    theories collectively? If one of these theories does not apply, then another one comes in;

43    if none of them applies, there must be one yet to be identified. At the same time, all the

44    aforementioned theories remain correct collectively!

45        No, I am not referring to astrology; I am not referring to some ancient

46    superstition. I am referring to an important field in the study of evolution; Haldane's rule

47    and the composite theory. The core of the composite theory – the dominance theory went

48    through just such episodes of acceptance, refutation and resurrection. The dominance

49    theory, together with the faster male theory, is the sub-theory of the so-called composite

50    theory. These theory and sub-theories together (thus "composite") have been declared to

51    be the 'correct' explanation of Haldane's rule by leading investigators in the field (Orr,

52    1997; Turelli, 1998).

53        Haldane's rule is a phenomenon that was first formulated in 1922 by JBS Haldane

54    through the examination of hybridization data in literature (Haldane, 1922): "When in the

55    $F_1$ offspring of two different animal races one sex is absent, rare, or sterile, that sex is the

56    heterozygous sex [heterogametic sex]." Haldane's rule is one of the most consistent

57    patterns in early speciation of sexually reproducing animals. It concerns a form of

58    postzygotic isolation frequently observed in early speciation: the pervasive occurrence of

59    sterility or inviability in $F_1$ hybrids of the heterogametic ($XY$ or $ZW$) sex than the

60    homogametic (*XX* or *ZZ*) sex in hybridization between closely related species or

61    subspecies. In mammals and in Drosophila (*XY* sex determination, males are *XY* and

62    females *XX*), the affected sex is male; in birds and in butterflies (*ZW* sex determination,

63    females are *ZW* and males *ZZ*), the affected sex is female. By appearance, Haldane's rule

64    is a phenomenon associated with the heterogamety of sex chromosomes. The rule has

65    been documented and confirmed repeatedly, including all major taxa of diecious animals

66    and plants (Coyne and Orr, 1989; Johnson, 2000; Laurie, 1997).

67        Haldane's rule is almost an obligative step during early speciation and imposes

68    one of the fundamental questions in speciation study: how postzygotic isolation evolves

69    during early speciation and why the heterogametic sex is much more vulnerable to hybrid

70    inferiority (sterility and inviability) than the homogametic sex. How heterogamety plays a

71    role in Haldane's rule is one of the most intriguing questions of speciation genetics.

72    Currently, the cause of Haldane's rule is claimed to be a "solved" problem. The

73    mainstream view believes that Haldane's rule is coincidentally/collectively caused by

74    multiple mechanisms. This so-called composite theory subdivides Haldane's rule and

75    invokes different explanations for different subdivisions (Turelli, 1998). Based on this

76    theory, Haldane's rule for sterility and inviability, male heterogamety and female

77    heterogamety belongs to separate subdivisions and require different explanations. It is

78    also believed that different explanatory theories operate in certain subdivisions

79    individually or collectively (Turelli and Orr, 2000). I am strongly opposed to the

80    composite theory and its major sub-theories. I will start by briefly reiterating the history

81    of the study.

82       In 1940s, Muller proposed that the epistatic recessive defects of loci linked to the

83    *X* chromosome caused imbalance of gene expression that leads to sterility/inviability in

84    hybrids. Such *X* linked recessiveness would not affect individuals in their native

85    population due to coevolved autosomal background that mask the defects. After

86    hybridization, however, a heterogametic $F_1$ hybrid carries only one *X* chromosome

87    (hemizygote), the recessive incompatibility would thus be expressed and cause sterility or

88    inviability in the heterogametic sex. A homogametic $F_1$ hybrid, on the other hand, carries

89    two *X* chromosomes (heterozygote); one of each from both parental populations, the

90    recessiveness of the *X* would be masked by the dominant allele on the other *X* and cause

91    no sterility or inviability. In this scenario, the *X* chromosomes and autosomes are all

92    heterozygous and Muller suggested that the recessive defects on the *X* would be balanced

93    out by the corresponding autosomes (Muller, 1940; Muller, 1942; Muller and Pontecorvo,

94    1942).

95       Muller's explanation was originally known as the *X*-autosome imbalance theory

96    (Muller, 1940; Muller, 1942; Muller and Pontecorvo, 1942), and later renamed as the

97    dominance theory to better describe its dominant/recessive nature (for consistency, I will

98    use the dominance theory throughout in the following). Until 1985, Muller's explanation

99    had been considered to be the general explanation as to how and why the heterogametic

100    $F_1$ hybrids are more susceptible to sterility and inviability (Laurie, 1997).

101       In 1985, Coyne published a monumental report of a Drosophila experiment that

102    negated Muller's dominance theory. In a hybridization experiment between *Drosophila*

103    *simulans* and its sibling species *D. sechellia* and *D.mauritania*, Coyne engineered female

104    hybrids that carried two identical *X* chromosomes from *D. simulans* with an otherwise $F_1$

105    genetic background where the male was sterile. Based on the dominance theory, this is a

106    scenario where the recessive imbalance between the $X$ chromosome and an autosome(s)

107    should lead to female sterility. However, the female $F_1$ that Coyne obtained were fertile.

108    The predicted recessive locus/loci on the homozygous $X$ chromosomes, expected to cause

109    female sterility based on the dominance theory, failed to cause sterility in these female $F_1$

110    hybrids, which carried two identical $X$ chromosomes of *D. simulans* (Coyne, 1985). This

111    classic study prompted an intense interest in seeking for an alternative explanation for

112    Haldane's rule.

113         The focus had been mainly on searching for an alternative genetic cause that

114    applied unitarily to all or most cases. However, these efforts were unsuccessful (Coyne,

115    1992). Exhaustive search and analysis of various possibilities failed to produce a single

116    unitary genetic mechanism of Haldane's rule that many were searching for. Besides the

117    dominance theory, the mechanisms ever proposed cover a wide array of cytogenetic

118    incompatibilities, *e.g.* chromosomal rearrangements (Haldane, 1932), dosage

119    compensation (Cline and Meyer, 1996), $X$-$Y$ incompatibilities (Heikkinen and Lumme,

120    1998; Muller, 1942), $Y$-autosomal incompatibilities (Heikkinen and Lumme, 1998;

121    Pantazidis and Zouros, 1988; Pantazidis et al., 1993), and meiotic drive (Frank, 1991;

122    Hurst and Pomiankowski, 1991). Another curious fact from these studies is that while

123    none of these mechanisms qualifies as the general genetic basis, many of them seem

124    perfectly applicable to some isolated cases. This puzzling situation came to an end in

125    1992.

126         In 1992, Wu proposed that Haldane's rule may not have a single genetic basis,

127    based on discrepancies in literature: two previous observations of hybrid inviability

128   supported Muller's dominance theory but Coyne's test on sterility clearly negated the

129   Muller's theory (Wu, 1992). Wu reasoned that Haldane's rule was possibly a composite

130   phenomenon caused by multiple mechanisms, and Haldane's rule for sterility and

131   inviability may belong to different entities that require different explanations. Based on

132   this notion, Wu compared the evolution rates of complete (or nearly complete) inviability

133   and sterility in Drosophila and mammals compiled previously by two other researchers,

134   and discovered that the evolution rate of hybrid sterility in Drosophila and mammals were

135   much faster than that of inviability (Wu, 1992).

136         In 1993, Wu and Davis elaborated further on the idea of Haldane's rule being a

137   composite phenomenon requiring different explanations. Wu and Davis provided a more

138   extensive literature examination to support the faster male theory. The arguments Wu and

139   Davis made were that: (1) genes causing sterility usually behave sex-dependently but

140   those causing hybrid inviability do not; (2) the cases of Haldane's rule for sterility

141   outnumbers those for inviability by more than 10-fold in Drosophila and mammals; and

142   (3) in Drosophila, genes causing hybrid male sterility greatly outnumber genes causing

143   male inviability, but mutagenesis experiments indicated that mutations affecting viability

144   outnumber those sterility. Therefore, BDM isolation causing sterility evolved much faster

145   than ones causing inviability and they believed that Haldane's rule for sterility was a

146   result of such evolutionary dynamics. Wu and Davis suggested that the dominance theory

147   remained to be the valid explanation for Haldane's rule for inviability. Also, Wu and

148   Davis suggested that the sterility component of Haldane's rule should be further

149   subdivided, and Haldane's rule for sterility in male heterogametic species and in female

150   heterogametic species may have different causes. The faster male theory was offered as a

151    general mechanism of Haldane's rule for sterility in taxa with *XY* sex determination (Wu

152    and Davis, 1993). In the same paper, Wu and Davis provided clear descriptions of the

153    composite theory: (1) Haldane's rule is a composite phenomenon that can be divided into

154    different subdivisions; and (2) Haldane's rule for sterility and inviability has different

155    causes (Wu and Davis, 1993).

156         The notion of Haldane's rule being a composite phenomenon and Haldane's rule

157    for sterility and inviability requiring different explanations were quickly embraced by

158    others. In 1993, Orr published a very similar experiment to Coyne's test of sterility, but

159    Orr tested inviability with the cross of *D. simulus* and *D. teissieri* (Orr, 1993a). The cross

160    of *D. simulus* and *D. teissieri* obeys Haldane's rule by $F_1$ male inviability. What he found

161    was that the homozygous *X* chromosomes of *D. simulus* did cause inviability in the

162    female hybrids in an otherwise $F_1$ male genetic background. Orr demonstrated that the

163    recessiveness that causes $F_1$ male inviability could cause $F_1$ female inviability. In Orr's

164    experiment, the supposed recessive defects did indeed caused inviability in the

165    engineered females, in contrast to what Coyne found in 1985 in a similar setting to test

166    sterility where the homozygous *X* chromosomes failed to cause sterility as the dominance

167    theory predicted. Orr's experiment convincingly demonstrated that recessive defects

168    could indeed cause Haldane's rule for inviability in the species pair of *D. simulus* and *D.*

169    *teissieri* (Orr, 1993a). In the same year, Orr provided a mathematical interpretation how a

170    partial recessive incompatibility could cause Haldane's rule for inviability (Orr, 1993b).

171    He declared that a modified version of the dominance theory could explain Haldane's

172    rule for inviability. In 1995, Turelli and Orr followed up with an additional mathematical

173    elaboration (Turelli and Orr, 1995). In 2000, Turelli and Orr further expanded the

174      applicability of this theory to hybrid sterility (Turelli and Orr, 2000). A third point was

175      also added into the composite theory by Orr and Turreli (2000), which was that different

176      mechanisms not only operate coincidentally, they also operate collectively to cause

177      Haldane's rule. These two papers (Turelli and Orr, 1995; Turelli and Orr, 2000) are

178      deemed to be the mathematical validation of the dominance theory.

179      The claim that because Haldane's rule has multiple genetic bases, it therefore has

180      multiple causes was never seriously challenged and extensively tested. Up to now,

181      investigators often equate the cause of Haldane's rule and the genetic bases of Haldane's

182      rule, and use the cause and the genetic bases of Haldane's rule interchangeably (Orr,

183      1993b; Turelli, 1998). It has become the dominant belief in the field that Haldane's rule

184      consists of multiple subdivisions that require different explanations (Orr, 1993a; Orr,

185      1997; Turelli, 1998; Wu, 1992; Wu and Davis, 1993). It was announced that the cause of

186      Haldane's rule is a solved problem, what remained to be done was "about the genes that

187      cause postzygotic isolation" and direct genetic analyses assessing *X*-linked recessivity in

188      hybrids (Turelli, 1998).

189      I disagree. My disagreement towards the composite theory is based on the

190      following two reasons. First, I believe that the founders of the composite theory are

191      confused about the genetic bases and the cause of Haldane's rule. From the very

192      beginning since the birth of Haldane's rule, the quest to search for the mechanism that

193      causes Haldane's rule had been mainly focusing on a unitary genetic mechanism that

194      offers a general explanation (Dobzhansky, 1937; Haldane, 1932; Muller, 1940; Muller,

195      1942; Muller and Pontecorvo, 1942). Based on the evidence before Coyne's study (1985),

196      Muller's dominance theory seemed to be a quite reasonable proposition – an *X*-autosome

197  imbalance with *X*-linked recessive defects could indeed cause $F_1$ sterility/inviability in

198  the heterogametic but not in the homogametic sex. However, there has been no evidence

199  whatsoever that has proven that it is only a genetic mechanism(s) should be the cause of

200  Haldane's rule. The genetic cause of inferiority does not necessarily have to be the *cause*

201  of Haldane's rule. Haldane's rule always has genetic bases of a certain form(s), but it is

202  quite different from Haldane's rule as a 'rule' caused by certain genetic cause. Wu's

203  analysis and Orr's critical experiment and mounting evidence from many authors

204  convincingly demonstrated that Haldane's rule has multiple genetic bases. The logic that

205  Haldane's rule has to be caused by multiple mechanisms because of the presence of

206  multiple genetic bases in different cases is simply a fallacy used to justify the founders'

207  own theories.

208      What is more interesting is that through such reasoning (Orr, 1993b; Turelli,

209  1998; Wu and Davis, 1993), heterogamety became a non-essential part of Haldane's rule.

210  At least in the faster male theory, it is the sex, not the heterogamety, to be considered to

211  be the cause of Haldane's rule. The fact that Haldane's rule has been such an amazing

212  natural phenomenon is largely because its association with heterogamety and wide

213  applicability in broad taxa. The heterogamety and wide applicability both disappeared

214  under the composite theory (well probably not totally disappeared, heterogamety is

215  invoked when needed such as in the dominance theory under some dubious

216  presumptions).

217      Even more curiously, while non-unitary genetic causes was used as the evidence

218  for multiple causes of Haldane's rule, a mechanism other than genetics, *i.e.* the faster

219  male theory – a theory about evolutionary dynamics, was first to be invoked (Wu, 1992;

220    Wu and Davis, 1993). The dominance theory, on the other hand, is a theory about the

221    cytogenetic bases. It would a big case to prove if one wanted to claim that these two

222    mechanisms (plus others) at different levels (The dominance theory is about cytogenetics

223    and the faster male theory is about population dynamics) could operate together and

224    produce Haldane's rule – such a strikingly consistency in nature across broad taxa. This

225    issue seems easily resolved by declaring Haldane's rule as a coincident caused by

226    multiple mechanisms. And that leads to my second point.

227        Second, the composite theory lacks testability. By declaring Haldane's rule as a

228    coincident caused by multiple mechanisms, the founders of the composite theory relieved

229    themselves from the heavy burden of proof. Nobody bothered to prove or convince others

230    why Haldane's rule has to be a coincident except for the previous failure of finding a

231    unitary genetic mechanism. The reasoning and analysis leading to the composite theory at

232    best provided some corroborating and circumstantial evidences. Corroborating and

233    circumstantial evidences would be everywhere if one looks for them (Popper, 1963). That

234    was exactly what the founders of the composite theory did when they formulate their

235    theories. The corroborating and circumstantial evidences for the composite theory include

236    the faster male theory and the dominance theory (Orr, 1993a; Orr, 1997; Turelli, 1998;

237    Wu, 1992; Wu and Davis, 1993), the two that need the composite theory to justify their

238    own righteousness.

239        I found this kind of reasoning and generalization rather troublesome. The

240    dominance theory was proposed by Muller as the general explanation of Haldane's rule.

241    Coyne's experiment (1885) convincingly dismissed the possibility of the dominance

242    theory as the general explanation by demonstrating that the same incompatibility causing

243    male sterility failed to cause female sterility in the otherwise same genetic background.

244    Even if one might agree with the view that Haldane's rule for sterility and for inviability

245    are indeed different entities and have different causes, Orr's results only prove the

246    existence of dominance effects that could cause $F_1$ hybrid inviability in one cross. It is far

247    from proving the dominance effects as the general cause of Haldane's rule for inviability!

248         Another troublesome point is that in the dominance theory Turreli and Orr

249    elaborated, the dominance/recessiveness relationship on $X$ alleles was devised as a special

250    case of postzygotic isolation known as BDM incompatibilities (Orr, 1996). Even if one

251    might agree with the view that Haldane's rule for inviability indeed had its own cause

252    that was different from other subdivisions, why would not then those other forms of $X$-

253    autosome BDM isolation, which do not have dominance/recessive defects on the $X$ but

254    could cause sex-biased inviability in $F_1$, evolve during speciation? Why must BDM

255    incompatibilities with $X$-linked partial recessive defects be the pervasive form of $X$-

256    autosome BDM isolation in causing sex-biased inviability during early speciation but

257    other forms becomes invisible? Without addressing these outstanding questions, how can

258    the dominance theory be "correct" in explaining Haldane's rule for inviability?

259         So far, the only other test for the composite theory outside of cytogenetics is the

260    test for the faster male theory, which cannot stand alone and is not a theory about

261    heterogamety. The faster male theory cannot adequately explain why homogametic male

262    traits did not evolve faster and produce the reversal Haldane's rule, *i.e.* $F_1$ homogametic

263    sterility, in female heterogametic species such as butterflies and birds. An *ad hoc*

264    presumption was again made: the sterility component of Haldane's rule might be further

265 subdivided, and Haldane's rule for sterility in male heterogametic species and in female

266 heterogametic species may have different causes (Wu and Davis, 1993).

267     With the declaration of Haldane's rule as a coincidence caused by multiple

268 mechanisms, the composite theory becomes a theory too good to be true – it has an

269 enormous explanatory power to apply to practically any case of Haldane's rule. By

270 adopting multiple alternating theories at different levels, none of which need to stand

271 alone to prove the case, the composite theory becomes unfalsifiable and irrefutable. If

272 Haldane's rule in a certain instance cannot be explained by the dominance theory, then it

273 might be explained by the faster male theory; if it cannot be explained by either, then it

274 must be explained by some other mechanisms, identified or yet to be identified. This is a

275 theory that never fails. With such approach, just about any phenomenon or puzzle in

276 nature can be explained or solved by a "composite" theory of some sort. What is the use

277 of such a bulletproof and invincible theory for the advancement of science, and for the

278 advancement of speciation genetics? Karl Popper once wrote: "Irrefutability is not a

279 virtue of a theory" (Popper, 1963). Is the composite theory really a scientific theory?

280     I challenge the founders and proponents of the composite theory to prove the

281 composite theory as a theory testable, refutable and falsifiable, rather than a theory as the

282 ultimate truth for explaining Haldane's rule.

283     In short, the composite theory was proposed to provide *ad hoc* presumptions to

284 justify the faster male theory and the dominance theory, as a consequence of the failure to

285 find a unitary genetic cause of Haldane's rule. The presumptions that the composite

286 theory represents have not be tested and validated. I believe the composite theory does

287 not possess testability, falsifiability, and refutability that a real scientific theory requires,

288 and should be abandoned!

289
290 **References:**
291